\documentclass[twocolumn]{aastex63}

\usepackage{graphbox,graphicx}
\usepackage{dcolumn}
\usepackage{bm}
\usepackage{amssymb}
\usepackage{epstopdf}
\usepackage{amsmath}
\usepackage{amsfonts}
\usepackage{color}
\usepackage{mathrsfs}
\usepackage{xspace}

\newcommand{\be}{\begin{equation}}
\newcommand{\ee}{\end{equation}}
\newcommand{\ba}{\begin{eqnarray}}
\newcommand{\ea}{\end{eqnarray}}

\newcommand{\barr}{\begin{array}}
\newcommand{\earr}{\end{array}}

\newcommand{\bigoh}{\mathcal{O}}

\def\DM{{\rm DM}}
\def\Dg{{\rm DM}_{\rm gal}}
\def\Di{{\rm DM}_{\rm IGM}}
\def\Dh{{\rm DM}_{\rm host}}
\def\Dhh{{\rm DM}_{\rm hh}}
\def\SNR{{\rm SNR}}

\def\Var{\mbox{Var}}
\def\pvalue{$p\mkern1.55mu$-value\xspace}
\def\pvalues{$p\mkern1.55mu$-values\xspace}
\def\Lvalues{$L\mkern1.55mu$-values\xspace}
\def\clfg{\smash{$C_\ell^{fg}$}}

\def\clff{\smash{$C_\ell^{ff}$}}
\def\pcc{pc\,cm$^{-3}$\xspace}

\begin{document}

\title{Statistical association between the candidate repeating FRB\,20200320A and a galaxy group}
\shorttitle{FRB\,20200320A host association}

\author[0000-0001-7694-6650]{Masoud~Rafiei-Ravandi}
  \affiliation{Department of Physics, McGill University, 3600 rue University, Montr\'eal, QC H3A 2T8, Canada}
  \affiliation{Trottier Space Institute, McGill University, 3550 rue University, Montr\'eal, QC H3A 2A7, Canada}
\author[0000-0002-2088-3125]{Kendrick~M.~Smith}
  \affiliation{Perimeter Institute for Theoretical Physics, 31 Caroline Street N, Waterloo, ON N25 2YL, Canada}
\author[0000-0002-2551-7554]{D.~Michilli}
  \affiliation{MIT Kavli Institute for Astrophysics and Space Research, Massachusetts Institute of Technology, 77 Massachusetts Ave, Cambridge, MA 02139, USA}
  \affiliation{Department of Physics, Massachusetts Institute of Technology, 77 Massachusetts Ave, Cambridge, MA 02139, USA}
\author[0000-0002-4795-697X]{Ziggy Pleunis}
  \affiliation{Dunlap Institute for Astronomy \& Astrophysics, University of Toronto, 50 St.~George Street, Toronto, ON M5S 3H4, Canada}
\author[0000-0002-3615-3514]{Mohit Bhardwaj}
  \affiliation{Department of Physics, Carnegie Mellon University, 5000 Forbes Avenue, Pittsburgh, 15213, PA, USA}
\author[0000-0001-7166-6422]{Matt Dobbs}
  \affiliation{Department of Physics, McGill University, 3600 rue University, Montr\'eal, QC H3A 2T8, Canada}
  \affiliation{Trottier Space Institute, McGill University, 3550 rue University, Montr\'eal, QC H3A 2A7, Canada}
\author[0000-0003-3734-8177]{Gwendolyn~M.~Eadie}
   \affiliation{Dunlap Institute for Astronomy \& Astrophysics, University of Toronto, 50 St.~George Street, Toronto, ON M5S 3H4, Canada}
   \affiliation{Department of Statistical Sciences, University of Toronto, 700 University Ave., Toronto, ON M5G 1Z5, Canada}
\author[0000-0001-8384-5049]{Emmanuel Fonseca}
  \affiliation{Department of Physics and Astronomy, West Virginia University, P.O. Box 6315, Morgantown, WV 26506, USA }
  \affiliation{Center for Gravitational Waves and Cosmology, West Virginia University, Chestnut Ridge Research Building, Morgantown, WV 26505, USA}
\author[0000-0002-3382-9558]{B.~M.~Gaensler}
  \affiliation{Dunlap Institute for Astronomy \& Astrophysics, University of Toronto, 50 St.~George Street, Toronto, ON M5S 3H4, Canada}
  \affiliation{David A.~Dunlap Department of Astronomy \& Astrophysics, University of Toronto, 50 St.~George Street, Toronto, ON M5S 3H4, Canada}
  \affiliation{Division of Physical and Biological Sciences, University of California Santa Cruz, Santa Cruz, CA 95064, USA}
\author[0000-0003-4810-7803]{Jane Kaczmarek}
  \affiliation{CSIRO Space \& Astronomy, Parkes Observatory, P.O. Box 276, Parkes NSW 2870, Australia}
  \affiliation{Department of Computer Science, Math, Physics, \& Statistics, University of British Columbia, Kelowna, BC V1V 1V7, Canada}
\author[0000-0001-9345-0307]{Victoria~M.~Kaspi}
  \affiliation{Department of Physics, McGill University, 3600 rue University, Montr\'eal, QC H3A 2T8, Canada}
  \affiliation{Trottier Space Institute, McGill University, 3550 rue University, Montr\'eal, QC H3A 2A7, Canada}
\author[0000-0002-4209-7408]{Calvin Leung}
  \affiliation{Department of Astronomy, University of California Berkeley, Berkeley, CA 94720, USA}
\author[0000-0001-7931-0607]{Dongzi Li}
  \affiliation{Cahill Center for Astronomy and Astrophysics, MC 249-17 California Institute of Technology, Pasadena CA 91125, USA}
\author[0000-0002-4279-6946]{Kiyoshi W.~Masui}
  \affiliation{MIT Kavli Institute for Astrophysics and Space Research, Massachusetts Institute of Technology, 77 Massachusetts Ave, Cambridge, MA 02139, USA}
  \affiliation{Department of Physics, Massachusetts Institute of Technology, 77 Massachusetts Ave, Cambridge, MA 02139, USA}
\author[0000-0002-8897-1973]{Ayush Pandhi}
  \affiliation{David A.~Dunlap Department of Astronomy \& Astrophysics, University of Toronto, 50 St.~George Street, Toronto, ON M5S 3H4, Canada}
  \affiliation{Dunlap Institute for Astronomy \& Astrophysics, University of Toronto, 50 St.~George Street, Toronto, ON M5S 3H4, Canada}
\author[0000-0002-8912-0732]{Aaron B.~Pearlman}
  \affiliation{Department of Physics, McGill University, 3600 rue University, Montr\'eal, QC H3A 2T8, Canada}
  \affiliation{Trottier Space Institute, McGill University, 3550 rue University, Montr\'eal, QC H3A 2A7, Canada}
\author[0000-0002-9822-8008]{Emily Petroff}
  \affiliation{Perimeter Institute for Theoretical Physics, 31 Caroline Street N, Waterloo, ON N25 2YL, Canada}
\author[0000-0003-1842-6096]{Mubdi Rahman}
  \affiliation{Sidrat Research, 124 Merton Street, Toronto, ON M4S 2Z2, Canada}
\author[0000-0002-7374-7119]{Paul Scholz}
  \affiliation{Department of Physics and Astronomy, York University, 4700 Keele Street, Toronto, ON MJ3 1P3, Canada}
  \affiliation{Dunlap Institute for Astronomy \& Astrophysics, University of Toronto, 50 St.~George Street, Toronto, ON M5S 3H4, Canada}
\author[0000-0002-9761-4353]{David C.~Stenning}
  \affiliation{Department of Statistics \& Actuarial Science, Simon Fraser University, 8888 University Dr, Burnaby, BC V5A 1S6, Canada}
\newcommand{\allacks}{
}

\correspondingauthor{Masoud Rafiei-Ravandi}
\email{masoud.rafiei-ravandi@mcgill.ca}

\begin{abstract}
We present results from angular cross-correlations between select samples of CHIME/FRB repeaters and
galaxies in three photometric galaxy surveys, which have shown correlations with the first CHIME/FRB catalog
containing repeating and nonrepeating sources: WISE$\times$SCOS, DESI-BGS, and DESI-LRG. We find a
statistically significant correlation (\pvalue $<0.001$, after accounting for look-elsewhere factors) between a sample
of repeaters with extragalactic dispersion measure $(\DM)>395$\,\pcc and WISE$\times$SCOS galaxies with redshift $z>0.275$. We
demonstrate that the correlation arises surprisingly because of a statistical association between FRB\,20200320A
(extragalactic $\DM\approx550$\,\pcc) and a galaxy group in the same dark matter halo at redshift $z\approx0.32$.
We estimate that the host halo, along with an intervening halo at redshift $z\approx0.12$, accounts for at least
$\sim30\%$ of the extragalactic DM. Our results strongly motivate incorporating galaxy group and cluster
catalogs into direct host association pipelines for FRBs with $\lesssim1'$ localization precision, effectively utilizing
the two-point information to constrain FRB properties such as their redshift and DM distributions.
In addition, we find marginal evidence for a negative correlation at 99.4\% CL
between a sample of repeating FRBs with baseband data (median extragalactic $\DM=354$\,\pcc)
and DESI-LRG galaxies with redshift \smash{$0.3\le z<0.45$}, suggesting that the repeaters might be
more prone than apparent nonrepeaters to propagation effects in FRB-galaxy correlations due to
intervening free electrons over angular scales $\sim0\fdg5$.
\end{abstract}

\keywords{Cosmology (343); High energy astrophysics (739); Large-scale structure of the universe
(902); Radio transient sources (2008)}

\section{Introduction}
\label{sec:introduction}
One of the most rapidly evolving fields of astrophysics is the study of extragalactic
fast radio bursts (FRBs). FRBs are highly energetic ($\sim$$10^{36-42}$\,erg) flashes of
radio waves of unknown origins \citep[for a recent review, see][]{Petroff:2022aa}.
In contrast to their intrinsic pulse widths that last for $\sim$milliseconds,
the FRB arrival time at Earth can be delayed, e.g., from $\sim$seconds to $\sim$minutes
across the 400--800\,MHz band, proportional to $\nu^{-2}$, where $\nu$ is the observed frequency.
This dispersion is a result of the interaction between propagating radio waves and intervening free electrons
from the source to observer. In the dispersion relation, the constant of proportionality is defined as the dispersion
measure $\DM\equiv\int n_e dx$, where $x$ is the distance and $n_e$ is the free electron column
density along the line of sight, broadly accounting for contributions from the FRB host ($\Dh$),
intergalactic medium ($\Di$), and Milky Way ($\Dg$) as well as other structures (e.g., intervening
galaxy groups and clusters).

The first FRB was unearthed serendipitously from archival data dating back to 2001
\citep{Lorimer:2007wn}. Since then, radio telescopes such as the Canadian Hydrogen
Intensity Mapping Experiment \citep[CHIME;][]{CHIME:2022}  have dedicated a large fraction of
their computational and human resources to detecting and analyzing these mysterious
signals. At the time of writing, about 40 FRBs had been localized
to their host galaxies \citep[see, e.g.,][]{Chatterjee:2017aa,Bannister:2019aa,Ravi:2019aa,
Bhandari:2020aa,Heintz:2020aa,Law:2020aa,Macquart:2020aa,Marcote:2020ts,Bhardwaj:2021ab,
Bhandari:2022aa,Bhandari:2023aa,Gordon:2023aa,Law:2023aa,Michilli:2023aa,Sharma:2023aa}.
FRB host associations are usually carried out through cross-matching
coordinates of single galaxies with FRBs. However,
depending on galaxy survey completeness limits, galaxy redshifts, FRB localization uncertainty,
and DM, an FRB may be found to be plausibly associated with just one or many galaxies
\citep[for a review on direct host associations, see][]{Eftekhari:2017aa}.
Enabled through its large field of view ($\sim$200\,sq.\,deg.), wide bandwidth (400--800\,MHz), and
highly optimized software, the CHIME Fast Radio Burst instrument \citep[CHIME/FRB;][]{Collaboration:2018aa}
detects a few FRBs per day, which over the production period of a few years ($\bigoh(10^3)$ FRBs)
has resulted in a wide range of FRB population studies. CHIME/FRB sources have localization
uncertainties of $\sim$10$'$ (real-time intensity beams) or $\lesssim\,$1$'$ (through saved
voltage data) that are generally not sufficient for per-object analyses such as cross-matching
FRBs with auxiliary catalogs of, e.g., spectroscopic redshift galaxies or X-ray sources \citep[for an
exception, see] [in which a CHIME/FRB source was localized robustly to the outskirts of M81]{Bhardwaj:2021aa}.
Thus far, CHIME/FRB population studies have relied on the large number of
sources with localization uncertainties of $\sim$10$'$. For instance, \cite{Josephy:2021ts}
showed that CHIME/FRB Catalog 1 \citep{Collaboration:2021aa} does not correlate spatially
with the Galactic plane. To date, $\sim$3\% of CHIME/FRB sources have emitted repeating bursts
sporadically or periodically over long time intervals of $\sim$days--months \citep{Collaboration:2023aa}.
Compared to apparent nonrepeaters (hereafter, for brevity, nonrepeaters), repeaters exhibit wider
pulse widths and narrower bandwidths \citep{Pleunis:2021ui}.
Using the observed DM distributions, \cite{Collaboration:2023aa} showed that the repeating and nonrepeating
FRB populations differ statistically at $\approx2.5-3\sigma$ levels. In this work, we compare the two
populations through their angular cross-correlations with photometric redshift
catalogs of galaxies \citep[for a comprehensive formalism of FRB-galaxy correlations,
see][]{Rafiei-Ravandi:2020aa}.

This is a follow-up analysis based on \cite{Rafiei-Ravandi:2021aa}, who reported statistically
significant (\pvalues$\sim10^{-4}$) correlations between unique FRB sources (i.e., single bursts for
repeaters as well as nonrepeaters) in CHIME/FRB Catalog 1 (median extragalactic
\DM~$\approx$~536\,\pcc) and galaxies in the WISE$\times$SCOS, DESI-BGS, and DESI-LRG
surveys. \cite{Rafiei-Ravandi:2021aa} showed that the strength and angular scale of FRB-galaxy
correlations are consistent with an order-one fraction of the Catalog 1 FRBs inhabiting the same dark matter halos
that host galaxies in the redshift range $0.3\lesssim z\lesssim 0.5$. In addition, \cite{Rafiei-Ravandi:2021aa}
found statistical evidence for a subpopulation
of FRBs with high $\Dh$ ($\sim$400\,\pcc), and showed that this could be explained through the
host halo DM ($\Dhh$) contribution for FRBs near the centers of massive
($\sim$$10^{14}\,M_\odot$) halos.

In this work, we present results from the FRB-galaxy correlation
analysis for a sample of 25 newly cataloged repeaters and 14 candidate repeating FRB sources
\citep{Collaboration:2023aa}
along with 18 previously published repeaters \citep{Collaboration:2019_R2, Collaboration:2019aa,
Fonseca:2020aa, Michilli:2023aa}, 41 of which have baseband localizations.
The CHIME/FRB baseband pipeline \citep{Michilli:2021tm} records raw voltage data
that allow for sky localizations $\lesssim1'$, enabling
cross-correlation studies at high multipoles $\ell\sim10^4$, where a large number of
harmonic modes can give rise to a large signal-to-noise ratio from two-point correlations between
FRBs and galaxies in the same dark matter halos, i.e., the one-halo
term in the FRB-galaxy cross power spectrum
\citep[see, e.g., Figures 2 and 8 in][respectively]{Rafiei-Ravandi:2021aa,Rafiei-Ravandi:2020aa}.

Statistical cross-correlations between FRB and galaxy catalogs have significant constraining
power over a wide range of spatial scales, including the two-halo, one-halo, and Poisson
terms \citep[see][]{Rafiei-Ravandi:2020aa}. For FRB catalogs with sub-arcsec localization precision,
the two-point information is fully recovered in direct host associations, which attempt
to trace FRBs back to their origins in single galaxies. In this work, we present an exceptional
example that demonstrates how direct associations with catalogs of galaxy groups or clusters can still
recover information for constraining, e.g., an FRB redshift in the absence of an exactly identified
host galaxy. FRB-galaxy group/cluster associations have been proposed in a few recent works.
For example, \cite{Connor:2022aa} cross-matched CHIME/FRB Catalog 1 with a galaxy group
catalog \citep{Kourkchi:2017aa} in order to place modest constraints on the baryonic content of the
circumgalactic medium (CGM) in the local Universe ($<$\,40\,Mpc). More recently, \cite{Connor:2023aa}
identified two FRB hosts using the DESI group/cluster catalog from \cite{Yang:2021aa}.
At the time of writing, there had not been any published report on directly cross-matching catalogs
of galaxy groups or clusters with roughly localized FRBs that have no associated host galaxies (e.g.,
a large fraction of CHIME/FRB sources) in order to probe FRB-galaxy associations on one-halo scales
and hence place constraints on FRB properties such as their redshift distribution.

We have organized this article as follows. In Section~\ref{sec:data}, we describe
the input catalogs for our cross-correlation pipeline, which is summarized in
Section~\ref{sec:pipeline-overview}. Then, we tabulate and discuss the results in
Sections~\ref{sec:results} and \ref{sec:discussion}, respectively. Unless
specified otherwise, all DM values are extragalactic; we subtract $\Dg$ using
the YMW16~\citep{Yao:2017aa} model throughout.
Following \cite{Rafiei-Ravandi:2021aa},
we ignore the Milky Way halo \citep[$29 \lesssim \DM_{\rm halo} \lesssim 111$
\pcc,][]{Cook:2023aa,Ravi:2023aa}. Our results are not sensitive to these model
assumptions. We adopt a flat $\Lambda$CDM cosmology with a Hubble expansion
rate $h=0.67$, matter abundance $\Omega_m=0.315$,
baryon abundance $\Omega_b=0.048$,
initial power spectrum amplitude $A_s=2.10\times10^{-9}$,
spectral index $n_s=0.965$,
neutrino mass $\sum_\nu m_\nu=0.06$~eV,
and cosmic microwave background (CMB) temperature $T_{\rm CMB}=2.726$~K\@,
which are identical to the assumptions of \cite{Rafiei-Ravandi:2021aa}, and consistent
with the Planck results of \cite{Planck_2018}.

\section{Data}
\label{sec:data}
In this work, we define three FRB samples based on CHIME/FRB Catalog 1 and
published repeaters\footnote{CHIME/FRB data products are available at
\url{http://www.chime-frb.ca/}.} with different localization precisions:
\begin{enumerate}
    \item {\it Nonrepeaters:} our first sample contains 457 unique sources, which includes all the
        apparently nonrepeating FRB sources from CHIME/FRB Catalog 1 \citep{Collaboration:2021aa}.
        Following \cite{Rafiei-Ravandi:2021aa}, we exclude the sidelobe FRB events 20190210D, 20190125B,
        and 20190202B. The sample of 457 was localized through real-time intensity beamforming (also known
        as header localization), which has, on average, a localization error of $\theta_f\sim10'$.\footnote{The
        CHIME/FRB beam transfer function can be approximated by a Gaussian with a FWHM $\theta_f$, which is
        defined as the FRB localization error throughout \citep[][]{Rafiei-Ravandi:2021aa}.}
    \item {\it Repeaters (baseband):} 41 CHIME/FRB repeating sources with published baseband localizations
        \citep{Collaboration:2019_R2, Collaboration:2019aa, Fonseca:2020aa, Michilli:2023aa,
        Collaboration:2023aa}. CHIME/FRB baseband localizations result in, on average,
        localization errors of $\theta_f\lesssim1'$.
    \item {\it Repeaters (all):} 52 sources, comprised of the ``repeaters (baseband)'' sample
    	mentioned above and 11 recently published repeating sources \citep{Collaboration:2023aa}.
	The positions of the latter were obtained by combining multiple header localizations
	($\theta_f\sim1'$).\footnote{Header localizations have confidence regions that can span over multiple
	disjoint islands (see, e.g., Figure~4 of \cite{Collaboration:2023aa}).}
\end{enumerate}

\begin{figure*}[hbt!]
\centerline{
    \hspace{0.005cm}
    \includegraphics[align=t,width=8.2cm]{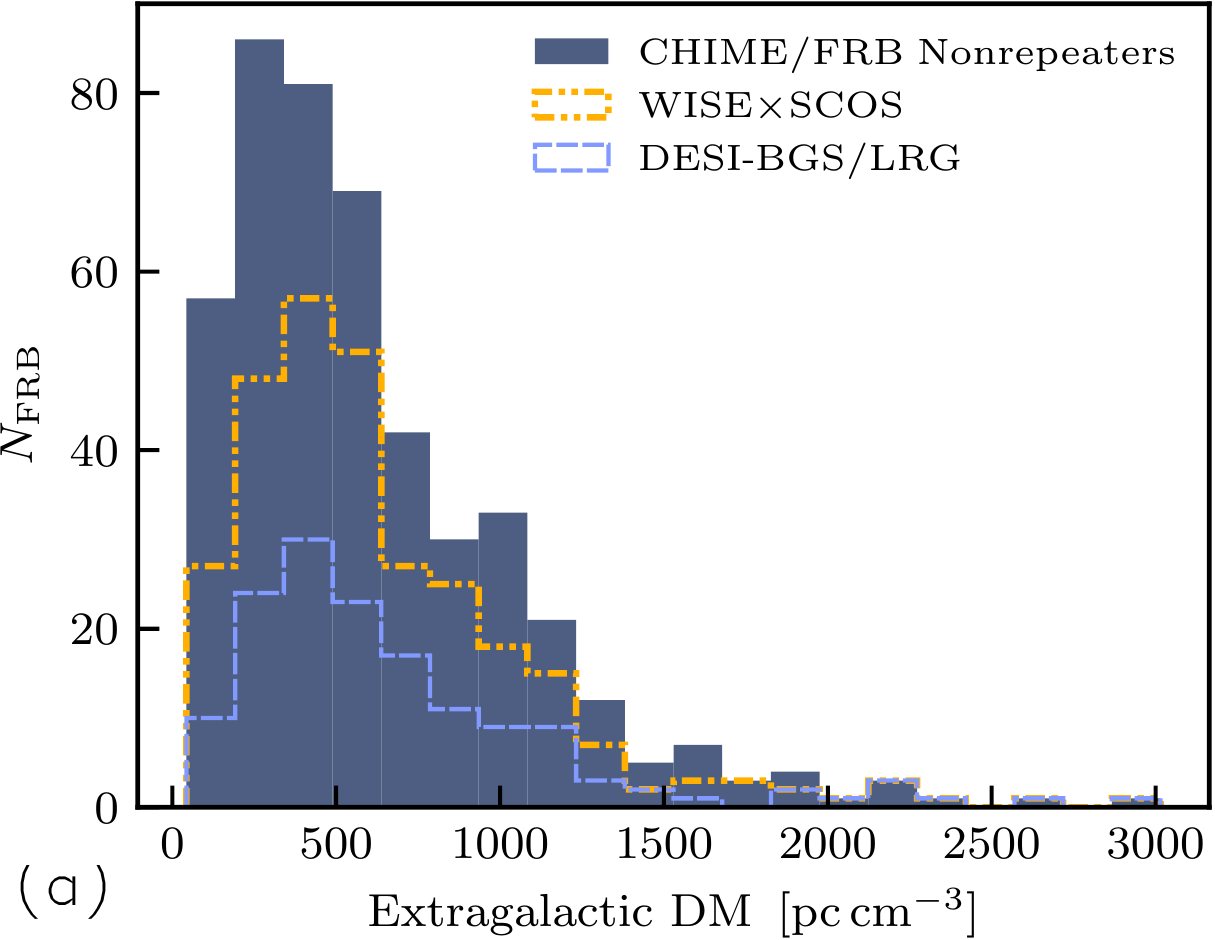}
    \hspace{0.7cm}
    \includegraphics[align=t,width=8.2cm]{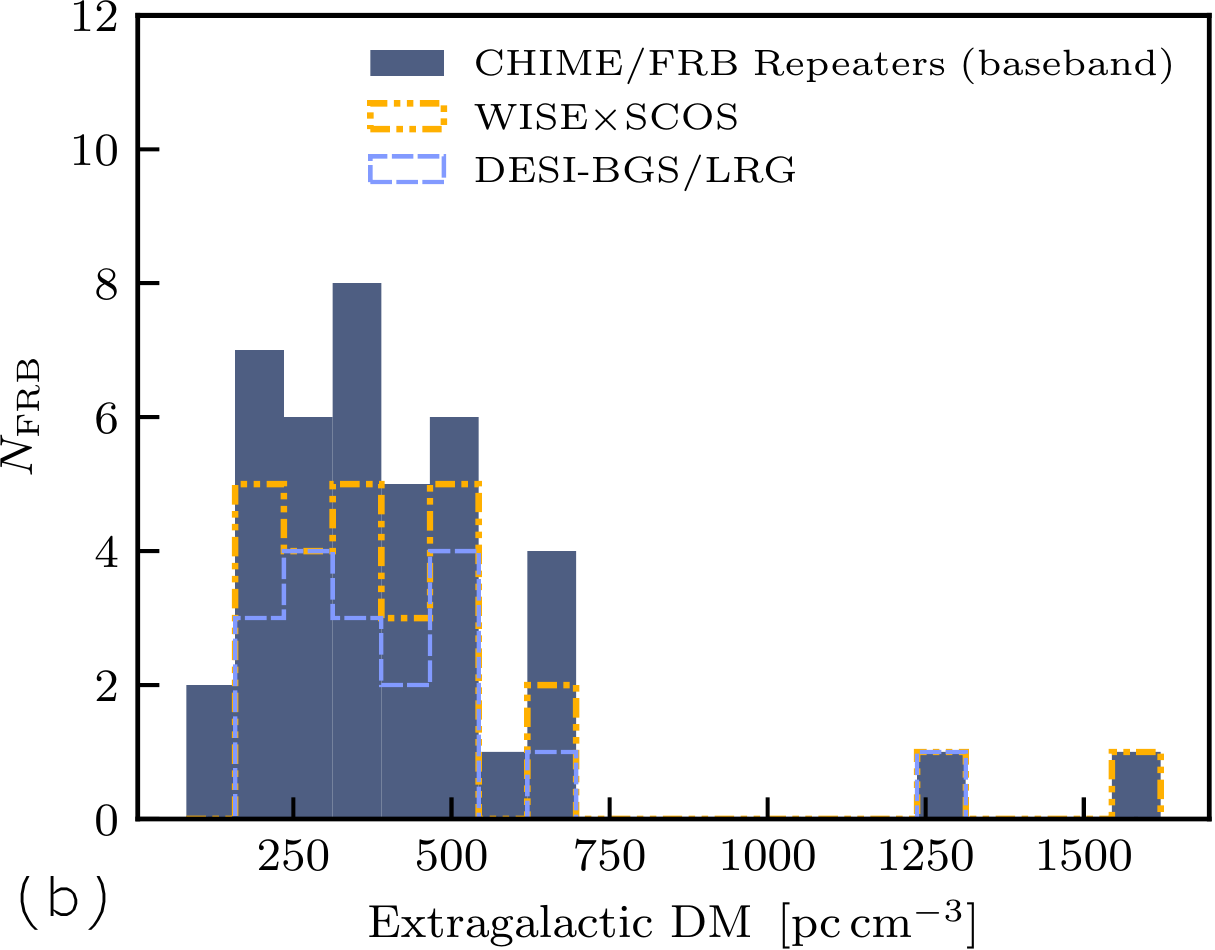}}
\vspace{0.32cm}
\hspace{0.15cm}
\centerline{
    \includegraphics[width=8.2cm]{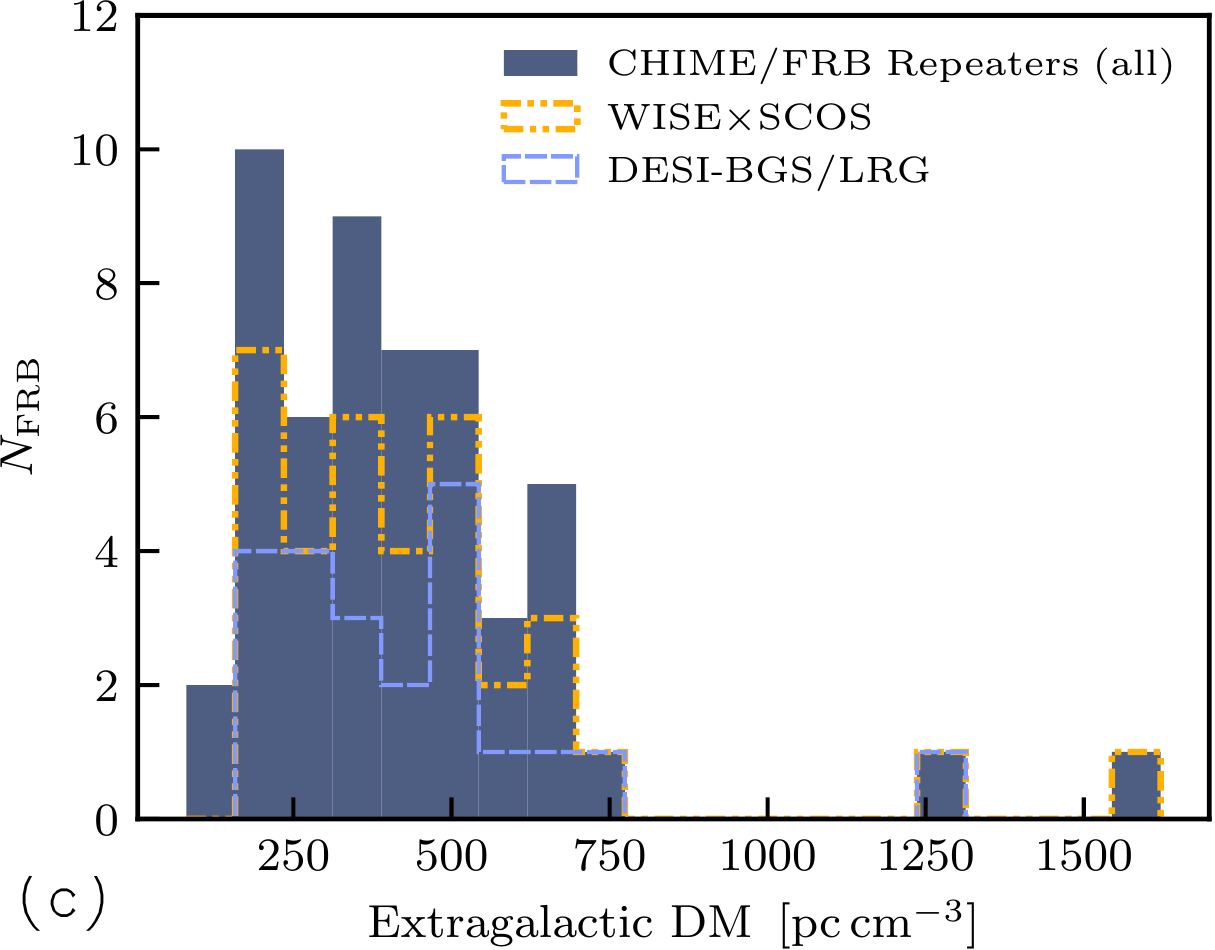}
    \hspace{0.4cm}
    \includegraphics[width=8.7cm]{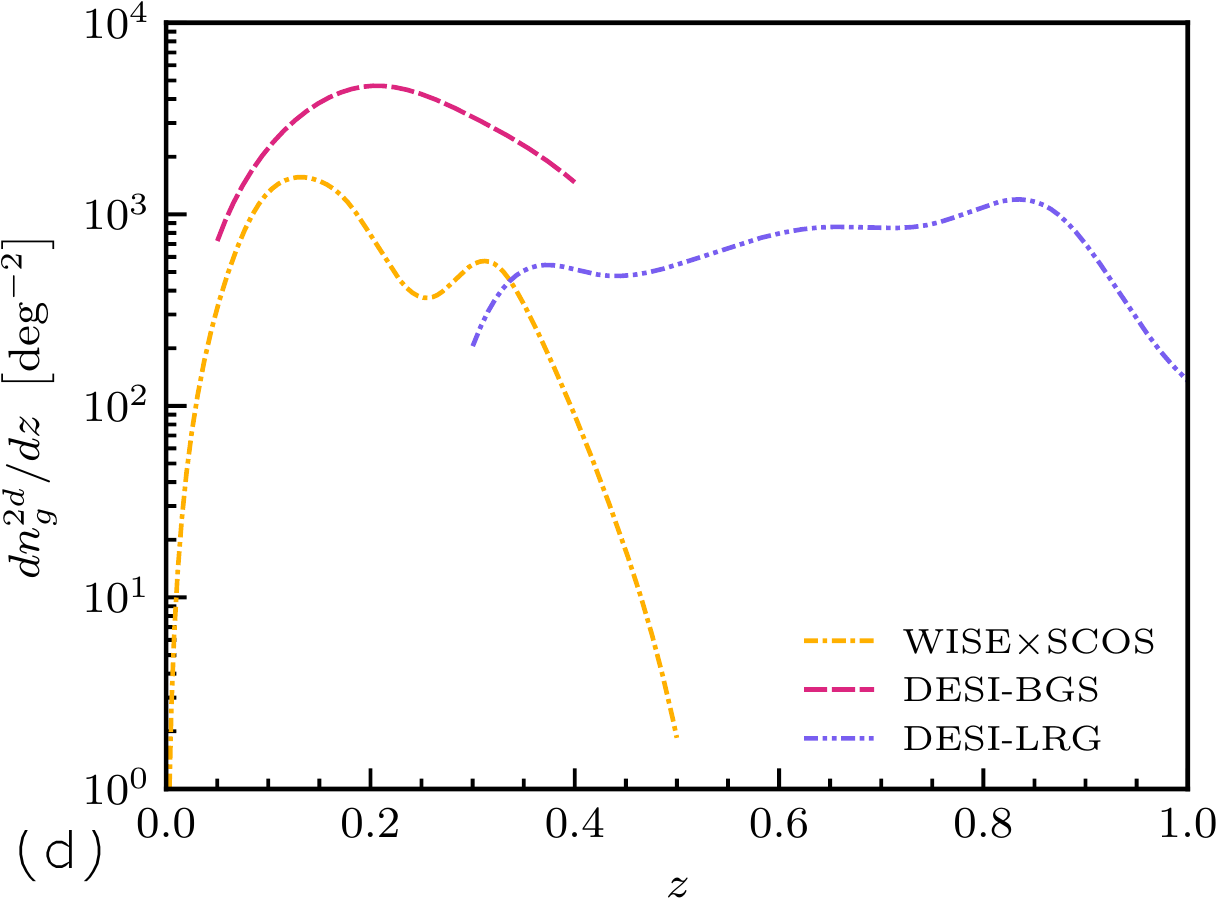}
}
    \caption{{\it Panels (a)--(c):} FRB DM distributions, including the subsets of FRBs that overlap with
    galaxy surveys for comparison. {\it Panel (d):} galaxy redshift distributions (see Table~\ref{tab:galaxy-surveys}).}
\label{fig:nfd}
\end{figure*}

Panels (a)--(c) of Figure~\ref{fig:nfd} show the DM distribution
for the three samples with median DM values 497 (a), 354 (b), and 372 (c)\,\pcc,
respectively.
\cite{Collaboration:2023aa} defined two distinct samples of repeaters based on the false positive
rate of FRB detections: 25 new repeaters (the ``gold'' sample) and 14 candidate repeating sources
(the ``silver'' sample). Because of the relatively small total number of repeating sources (compared
to nonrepeaters), we combine and use both samples along with other published repeaters throughout
this cross-correlation analysis. In addition, we simulate mock FRB catalogs independently for each FRB
sample listed above. Mock FRB catalogs have randomized right ascensions, but the same total number
of FRBs ($N_{\rm FRB}$) and other parameters, e.g., declinations, as in the real samples.

For galaxies, we select the three galaxy surveys WISE$\times$SCOS \citep{Bilicki:2016aa,Krakowski:2016aa},
DESI-BGS, and DESI-LRG \citep{Zhou:2020ab}, which correlate significantly
with CHIME/FRB Catalog 1 \citep{Rafiei-Ravandi:2021aa}. We adopt exactly the same galaxy selection cuts
\citep{Krakowski:2016aa, Ruiz-Macias:2020aa, Zhou:2020ac} and sky masks as adopted by
\cite{Rafiei-Ravandi:2021aa}. In Table~\ref{tab:galaxy-surveys}, we summarize the relevant properties of these
surveys after applying the selection cuts.
We note that only a fraction of the FRB samples are unmasked by the galaxy survey footprints (see, e.g.,
the dashed lines in panels (a)--(c) of Figure~\ref{fig:nfd}; the median DM values of nonrepeaters and repeaters
(all) change to 538 and 395\,\pcc after accounting for the WISE$\times$SCOS survey footprint, respectively).
Panel (d) of Figure~\ref{fig:nfd} shows the redshift distributions of these
galaxy surveys ($z\sim0.5$), which are desirable for probing statistical correlations with the CHIME/FRB
samples (extragalactic $\DM\sim500$\,\pcc).\\

\begin{table}
\begin{center}
\begin{tabular}{@{\hskip 0.0cm}c@{\hskip -0.4cm}ccc@{\hskip 0.0cm}}
\hline\hline
    Survey & WISE$\times$SCOS & DESI-BGS & DESI-LRG \\
    \hline
    $f_{\rm sky}$ & 0.638 & 0.118 & 0.118 \\
    $[z_{\rm min},z_{\rm max}]$ & [0.0,\,0.5] & [0.05,\,0.4] & [0.3,\,1.0] \\
    $z_{\rm med}$ & 0.16 & 0.22 & 0.69 \\
    $N_{\rm gal}$ & 6,931,441 & 5,304,153 & 2,331,043 \\
    \hline
    $N_{\rm FRB}$ & & & \\
    Nonrepeaters & 292 & 168 & 168 \\
    Repeaters\,(baseband) & 26 & 18 & 18 \\
    Repeaters\,(all) & 35 & 22 & 22 \\
    Catalog 1 & 310 & 183 & 183 \\ \hline\hline
\end{tabular}
\end{center}
    \caption{\label{tab:galaxy-surveys}Galaxy survey parameters: sky fraction $f_{\rm sky}$ \citep[not
    accounting for CHIME/FRB coverage; see][]{Rafiei-Ravandi:2021aa}, redshift range $[z_{\rm min},
    z_{\rm max}]$, median redshift $z_{\rm med}$, total number of unmasked galaxies $N_{\rm gal}$,
    and total number of FRBs $N_{\rm FRB}$ for different samples overlapping the survey.}
\end{table}

\section{Pipeline overview}
\label{sec:pipeline-overview}
\begin{figure*}[hbt]
\centerline{
    \includegraphics[align=t,width=8.6cm]{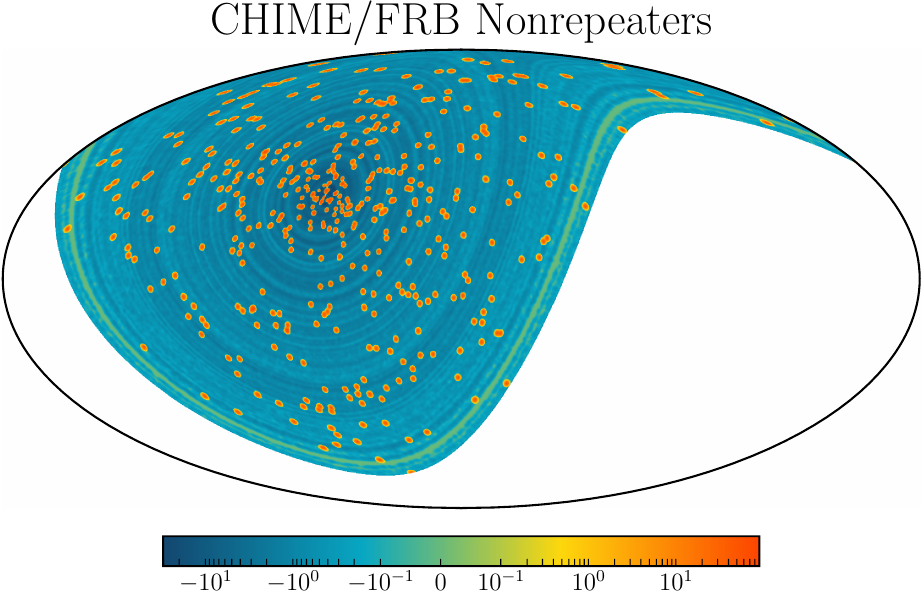}
    \hspace{0.8996153846153847cm}
    \includegraphics[align=t,width=7.6303846153846155cm]{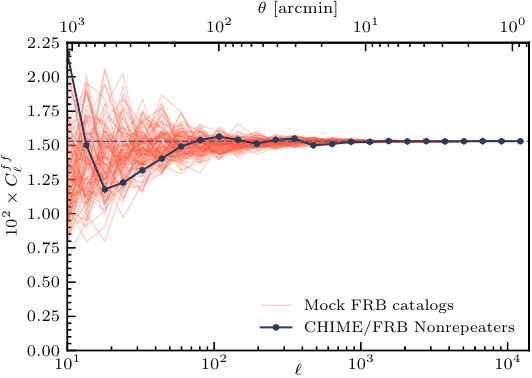}}
\vspace{0.75cm}
\centerline{
    \includegraphics[align=t,width=8.6cm]{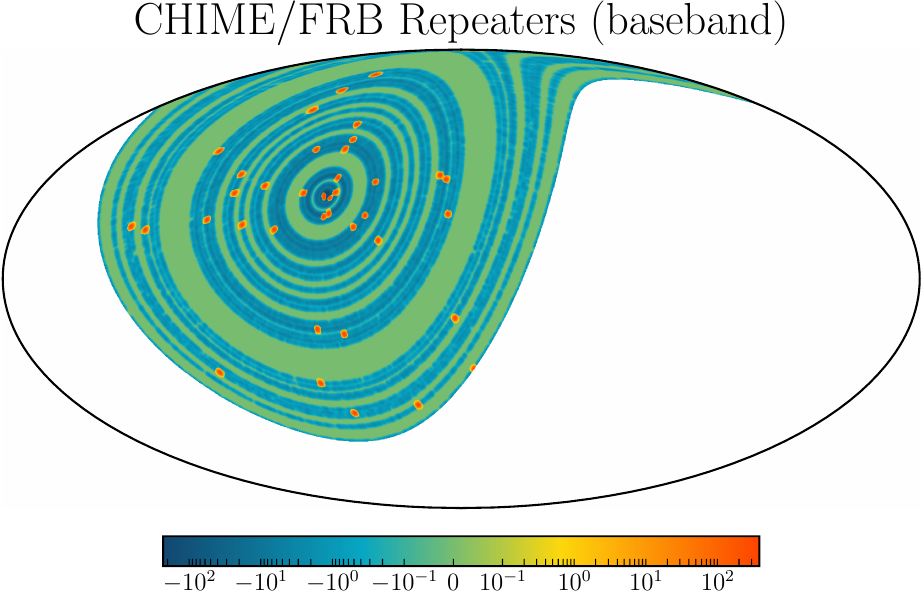}
    \hspace{0.75cm}
    \includegraphics[align=t,width=7.78cm]{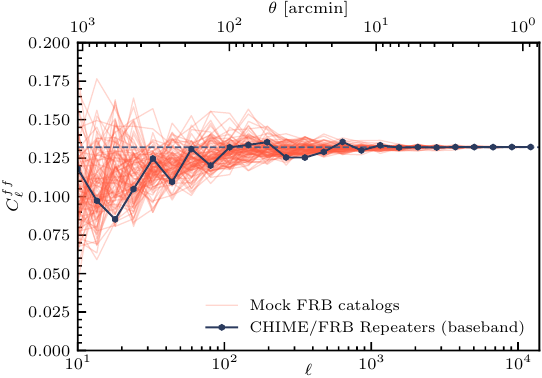}
}
\vspace{0.75cm}
\centerline{
    \includegraphics[align=t,width=8.6cm]{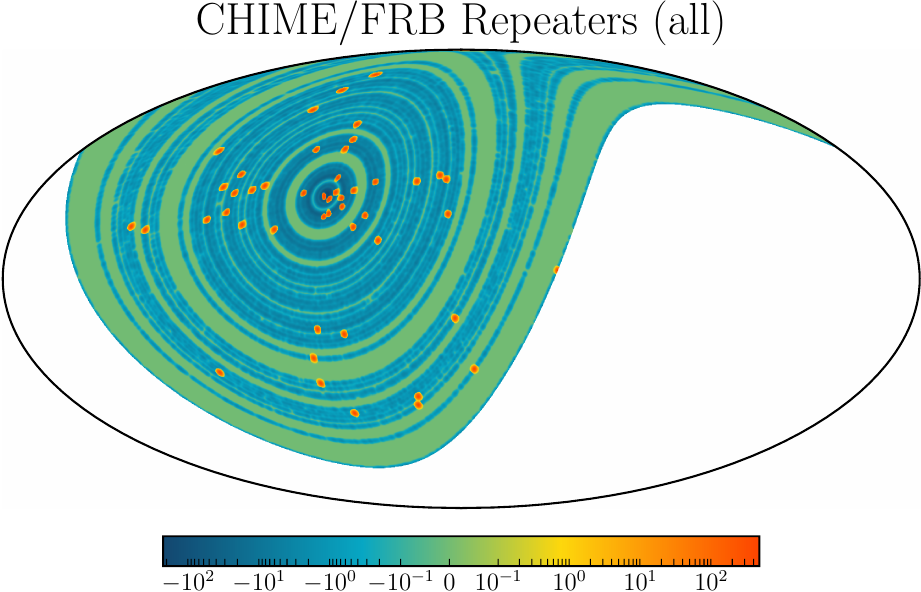}
    \hspace{0.75cm}
    \includegraphics[align=t,width=7.78cm]{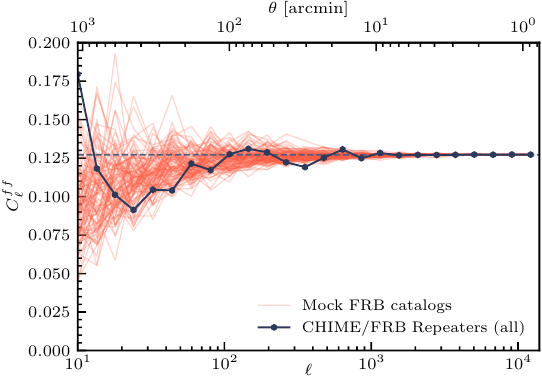}
\vspace{0.175cm}
}
    \caption{{\it Left column:} FRB overdensity maps, after incorporating random FRB
    catalogs, in Mollweide projection centered on
    Galactic longitude $l=180\degr$ in the Galactic coordinate system. Color bars
    indicate the full range of weighted density variations, including spurious fluctuations
    due to the survey geometry, which are modeled through ``random" FRBs that encircle the
    north celestial pole \citep[see][]{Rafiei-Ravandi:2021aa}.
    Each FRB (orange point) contributes $1/(n_f^{2d}
    \Omega_{\rm pix})$ to a pixel, where $n_f^{2d}$ is the 2D number density and
    $\Omega_{\rm pix}$ is the pixel area. {\it Right column:} angular auto power spectrum
    \clff for the three FRB samples on the left. Mock FRB catalogs (100 shown in each panel)
    trace the spatial distribution of real samples over a wide range of angular scales.
    Throughout, we assume Poisson-noise dominated FRB fields with \smash{$C_\ell^{ff}
    \approx 1/n_f^{2d}$} (dashed line) for template scales ($L$) corresponding to
    $315 \le \ell \le 13960$.}
\label{fig:overdensity-clff}
\end{figure*}

Our cross-correlation pipeline is described by \cite{Rafiei-Ravandi:2021aa}. Using the same
framework \citep{Rafiei-Ravandi:2023frbx},\footnote{FRBX: Tools for simulating, forecasting, and
analyzing statistical cross-correlations between FRBs and other cosmological sources
are available at \url{https://github.com/mrafieir/frbx}.} we generate high-resolution
HEALPix~\citep{Gorski:2005aa} overdensity maps ($N_{\rm side}=8192$) from the FRB ($f$)
and galaxy ($g$) samples \citep[see Figure~\ref{fig:overdensity-clff} here and
Figure~3 of][respectively]{Rafiei-Ravandi:2021aa}. Then, we estimate the amplitude of the
angular cross power spectrum \clfg, as a function of multipole (angular wavenumber) $\ell$,
to a maximum multipole of $\ell_{\rm max}=14000$. We assume that the power spectrum for
the one-halo term follows a template of the form \smash{$C_\ell^{fg(1h)} = \alpha e^{-\ell^2/L^2}$},
where the amplitude $\alpha$ (a free parameter) captures the one-halo term while the varying
template scale $L$ controls the high-$\ell$ suppression, owing to the CHIME/FRB
beam transfer function and FRB-galaxy displacements in the same dark matter halos.

Under the null hypothesis, the total one-halo term would be zero. We estimate the statistical significance
of a positive correlation (alternate hypothesis) by computing the maximum of local signal-to-noise
ratios $\SNR_{L,z} = \hat\alpha_{L,z} / \Var(\hat\alpha_{L,z})^{1/2}$ over the 2D parameter space of
template scales ($L$) and redshift endpoints ($z$) for the data and 1000 mock FRB catalogs.\footnote{The variance
$\Var(\hat\alpha_{L,z})$ is estimated from the ensemble of mock FRB catalogs.} Finally, we compute a global
\pvalue\ based on the fraction of mocks with the maximum local significance greater than or equal to the maximum
local significance of data: \smash{$\max_{L,z}^{\rm (mock)}\SNR_{L,z} \ge \max_{L,z}^{\rm (data)}\SNR_{L,z}$}.
This procedure fully accounts for the look-elsewhere effect that may arise because of our specific
choices in the search space ($L,z$). In this work, we also discuss the statistical significance of a
negative correlation (alternate hypothesis), which is computed through the same procedure as
described above, except that we replace all ``max'' operations with ``min''. Subsequently, the global \pvalue\
of a negative correlation is reported as the fraction of mocks with the minimum local significance
\smash{$\min_{L,z}^{\rm (mock)}\SNR_{L,z} \le \min_{L,z}^{\rm (data)}\SNR_{L,z}$}.
Throughout, we consider \pvalues\ of $\sim0.005$ and $\le0.001$ to be marginal
and significant evidence for the FRB-galaxy correlation, respectively.

As described extensively by \cite{Rafiei-Ravandi:2021aa}, the angular cross power
spectrum \clfg\ and hence all its associated statistics, including the global \pvalues, can be computed
for \emph{binned} catalogs. For instance, in Section~\ref{sec:results} of this work, we bin the repeaters (all)
sample by median DM in order to pin down the DM dependence of a significant correlation with WISE$\times$SCOS
galaxies that are binned by redshift endpoints. In Appendix A of \cite{Rafiei-Ravandi:2021aa}, we
showed that $L_{\rm max}=1396$ ($N_{\rm side}=2048$, $\ell_{\rm max}=2000$) is an appropriate choice for
CHIME/FRB Catalog 1 with header localizations. Assuming that baseband localization errors are 10 times
smaller than header localization errors, we extend the search over template scales out to
$L_{\rm max}=13960$, corresponding to the angular scale $\theta=\pi/L_{\rm max}=0\farcm77$ in this work.

\section{Results}
\label{sec:results}
\begin{figure*}[hbt!]
\centerline{
    \includegraphics[width=8.34051724137931cm]{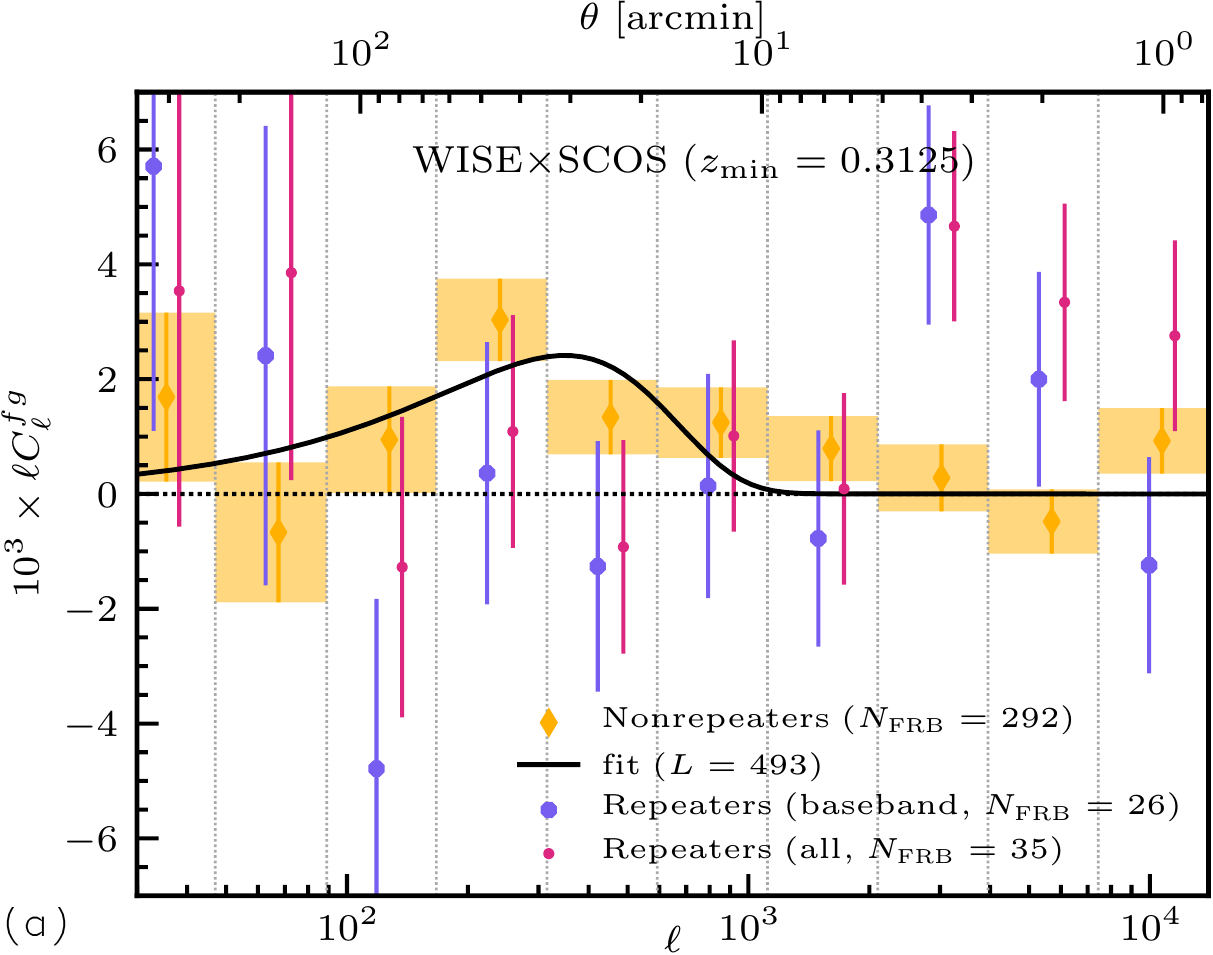}
    \hspace{0.4cm}
    \includegraphics[width=8.6cm]{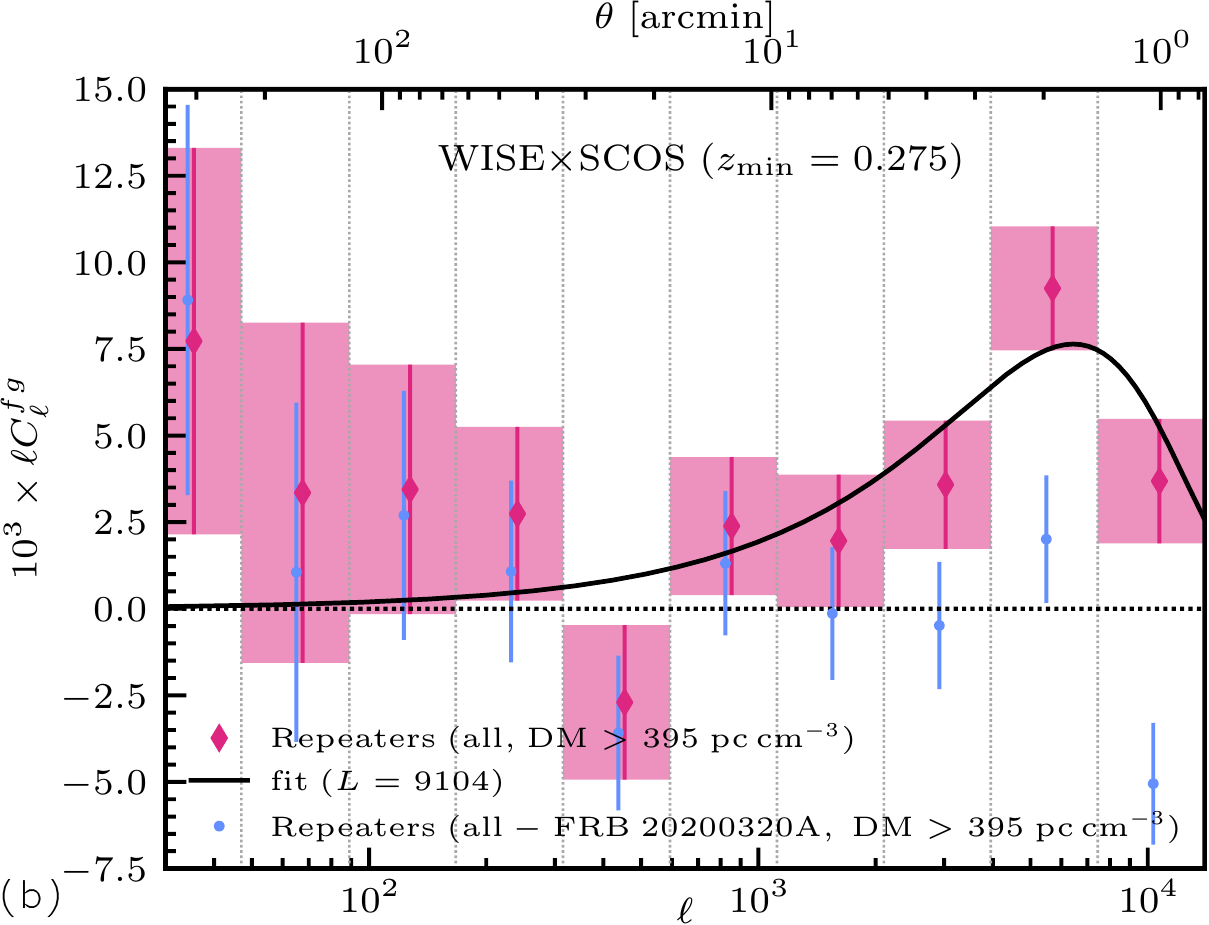}}
\vspace{0.4cm}
\centerline{
    \includegraphics[width=8.34051724137931cm]{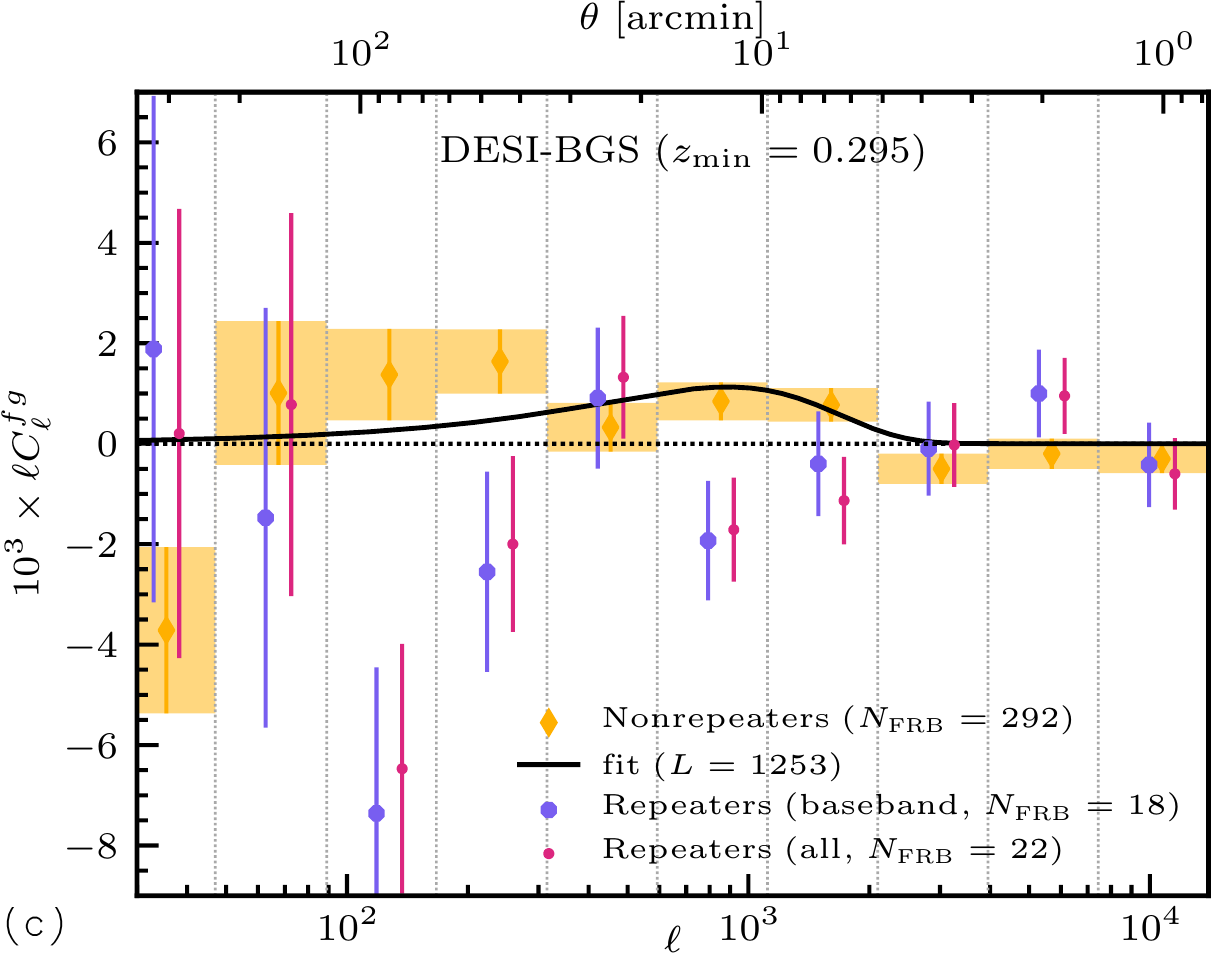}
    \hspace{0.2517241379310349cm}
    \includegraphics[width=8.748275862068965cm]{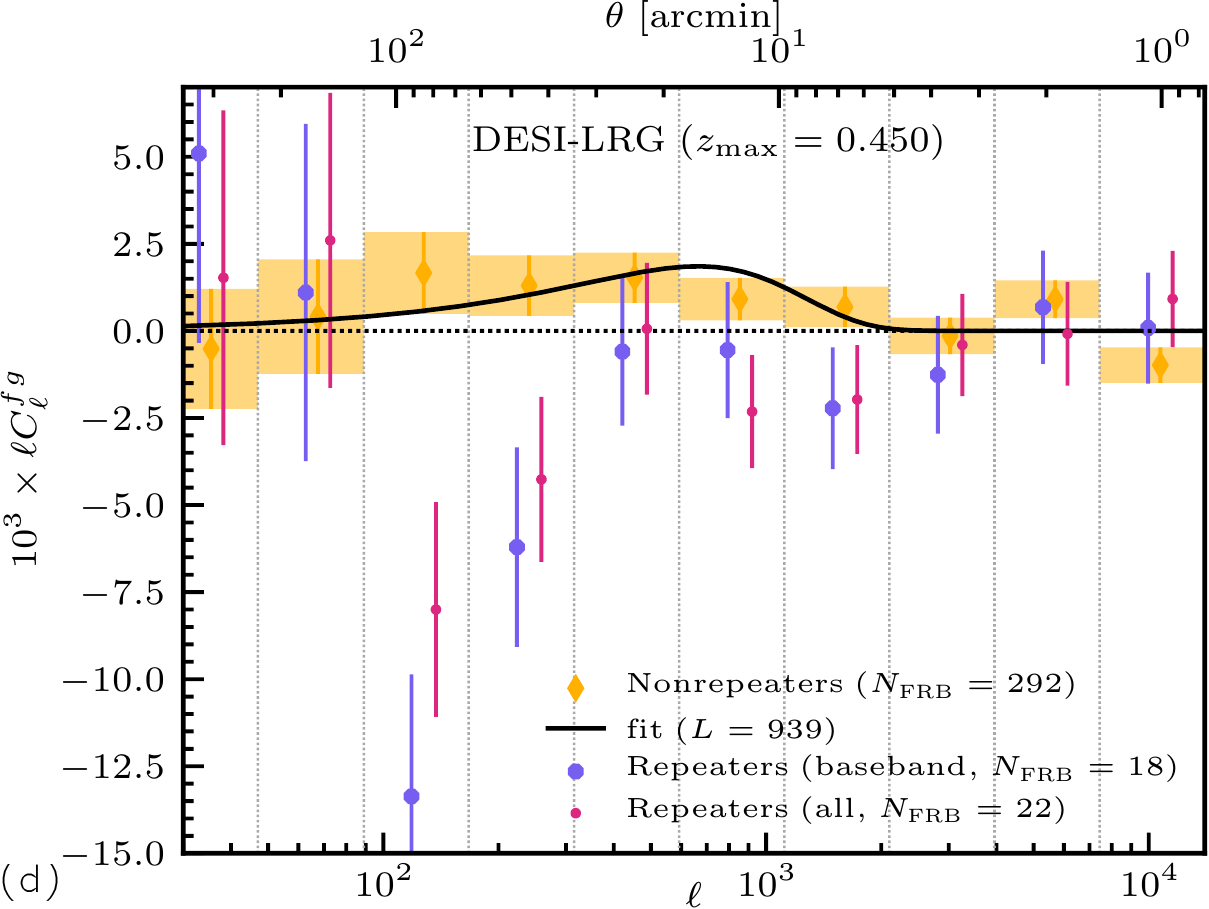}}
    \caption{Angular cross power spectrum \clfg\ for galaxy redshift bins (panels (a) through (d)) in which the
    ``nonrepeaters'' correlation with galaxies are maximized.
    The displayed ``fit" (solid black line) is the best-fit template of the form
    \smash{$C_\ell^{fg} = \alpha e^{-\ell^2/L^2}$}\,(fixed $L$) for
    the underlying cross power spectrum that is shown as shaded bandpowers.
    Error bars indicate 1$\sigma$ deviations based on 1000
    Monte Carlo realizations. Note the large error bars ($\propto N_{\rm FRB}^{-1/2}$,
    assuming \smash{$C_\ell^{fg} \ll (C_\ell^{ff} C_\ell^{gg})^{1/2}$}) for small
    FRB samples. In panel (b), the positive correlation between the DM-binned
    ``repeaters (all)'' sample and redshift-binned WISE$\times$SCOS galaxies is dominated by FRB\,20200320A.
    In panel (d), the negative correlations are either marginal or totally insignificant after accounting for
    look-elsewhere factors (see Section~\ref{sec:results}).}
    \label{fig:clfg}
\end{figure*}

In this section, we compare the repeating and nonrepeating FRB samples through their correlations
with galaxy catalogs. In addition, we report a significant correlation between a sample of repeaters
and WISE$\times$SCOS galaxies. We show that the correlation is dominated by a single FRB source.
Finally, we reproduce results for CHIME/FRB Catalog 1 with the updated pipeline parameters in this work.

Figure~\ref{fig:clfg} shows the FRB-galaxy cross power spectra for the repeaters and nonrepeaters.
We find that deviations in bandpowers (averaged power spectra in nonoverlapping $\ell$-bins)
are either consistent with each other within 1$\sigma$ (WISE$\times$SCOS
and DESI-LRG) or inconclusive (DESI-BGS) on scales of $\ell\sim1000$, where CHIME/FRB
Catalog 1 (along with the ``nonrepeaters,'' which make up the majority of the catalog) shows significant
positive correlations with the same galaxy surveys \citep[see Figure 7 of][]{Rafiei-Ravandi:2021aa}.
Focusing on the repeaters in panel (b) of Figure~\ref{fig:clfg}, we find a statistically significant
positive correlation between ``repeaters (all)'' with $\DM>395$\,\pcc (median extragalactic DM of the sample)
and galaxies with redshift $z>0.275$ in the WISE$\times$SCOS survey. The correlation has a local significance
with \smash{$\SNR_{\rm max}^{\rm (data)}=6.67$}. After accounting for look-elsewhere factors, we obtain
a global significance with a \pvalue$<0.001$ (see Table~\ref{tab:pvals}) for this positive correlation. In other words,
all the mock ``repeaters (all)'' catalogs with $\DM>395$\,\pcc correlate less significantly with WISE$\times$SCOS
galaxies over the search space of template scales and minimum redshift endpoints.

\begin{table*}[hbt!]
\begin{center}
\begin{tabular}{@{\hskip 0cm}cccccc@{\hskip -0.001cm}}
\hline\hline
    Sample & Repeaters & Repeaters & Repeaters & Nonrepeaters & Catalog 1 \\
    & (baseband) & (all) & (all$-$FRB\,20200320A) & & \\
\hline
    WISE$\times$SCOS & 0.278 & 0.007 & & $<0.001$ & $<0.001$\\
    & ($L$=1853, & ($L$=9750, & $\dots$ & ($L$=493, & ($L$=613,\\
    & $z_{\rm min}$=0.1375)& $z_{\rm min}$=0.275)& & $z_{\rm min}$=0.3125) & $z_{\rm min}$=0.3125)\\
    \\
    WISE$\times$SCOS & 0.373 & 0.470 & & 0.015 & 0.012 \\
    ($z_{\rm min}$=0.275) & ($L$=13960, & ($L$=13960, & $\dots$ & ($L$=954, & ($L$=989,\\
    & $\DM<395$\,\pcc)& $\DM<395$\,\pcc) & & $\DM<538$\,\pcc) & $\DM<535$\,\pcc)\\
    \\
    WISE$\times$SCOS & 0.463 & $<0.001$ & 0.757 & 0.012 & 0.035 \\
    ($z_{\rm min}$=0.275) & ($L$=5051, & ($L$=9104, & ($L$=5374, & ($L$=349, & ($L$=315,\\
    & $\DM>395$\,\pcc)& $\DM>395$\,\pcc) & $\DM>395$\,\pcc) & $\DM>538$\,\pcc) & $\DM>535$\,\pcc)\\
    \\
    DESI-BGS & 0.204 & 0.291 & & 0.006 & 0.001\\
    & ($L$=13636, & ($L$=9414, & $\dots$ & ($L$=1253, & ($L$=1355,\\
    & $z_{\rm min}$=0.050)& $z_{\rm min}$=0.050)& & $z_{\rm min}$=0.295) & $z_{\rm min}$=0.295)\\
    \\
    DESI-LRG & 0.545 & 0.506 & & 0.011 & 0.001\\
    & ($L$=1069, & ($L$=950, & $\dots$ & ($L$=939, & ($L$=1078,\\
    & $z_{\rm max}$=0.712)& $z_{\rm max}$=0.712)& & $z_{\rm max}$=0.45) & $z_{\rm max}$=0.485)\\ \hline\hline
\end{tabular}
\end{center}
    \caption{\label{tab:pvals}Statistical significance (\pvalue) of positive correlation between
    FRBs and galaxies after accounting for look-elsewhere factors in template scale ($L$)
    and redshift endpoint ($z_{\rm min}$ or $z_{\rm max}$). Rows and columns correspond
    to galaxy and FRB samples, respectively. The \Lvalues are characteristic scales for
    which \pvalues\ are presented. In the second and third rows, we select galaxies with the minimum
    redshift $z_{\rm min}=0.275$ (significant correlation with the unbinned ``repeaters (all)'' sample)
    and bin the FRB samples by DM. The DM cuts are median values after accounting for galaxy survey
    footprints (see Section~\ref{sec:data}).}
\end{table*}

\begin{figure}[hbt!]
\hspace{-0.07cm}\centerline{
    \includegraphics[width=8.46cm]{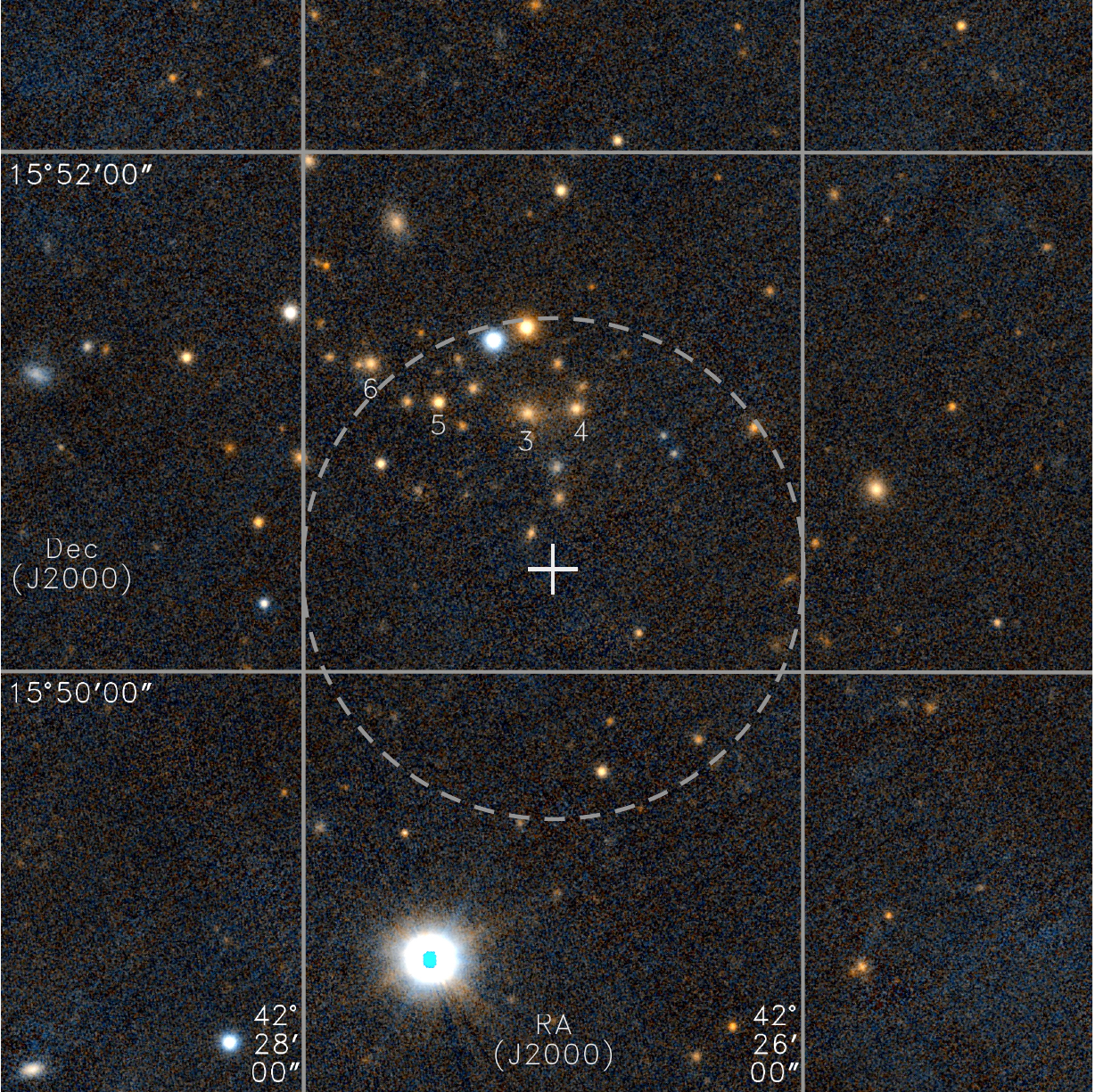}}
    \caption{Pan-STARRS DR1 color image \citep{Chambers:2016aa, Baumann:2022aa}
    of the vicinity of the candidate repeating FRB\,20200320A. The dashed line encircles objects
    within $1'$ from the center, where the FRB is localized. Using the cross-correlation pipeline,
    we find a statistical association between the FRB and four WISE$\times$SCOS
    galaxies, labeled \#3--6 underneath each object, presumably all in the same dark matter
    halo at $z\approx0.32$ (see Table~\ref{tab:wisexscos}).}
\label{fig:frb20200320a}
\end{figure}

Through visual inspection and pipeline reruns, we identify FRB\,20200320A in the vicinity of four galaxies
($z\sim0.3$) as the only source of this correlation (see Figure~\ref{fig:frb20200320a} and Table~\ref{tab:wisexscos}).
FRB\,20200320A is a candidate (``silver'') repeater with a signal-to-noise ratio of 284.8 and observed
inverse-variance-weighted average DM of 593.524(2)\,\pcc from the sample in \cite{Collaboration:2023aa}.
This candidate has two associated bursts that overlap within $1\sigma$ localization errors: FRB\,20200320A
(the first detection) and FRB\,20201105A (the second burst with a signal-to-noise ratio of 9.3 and observed DM of
581.408(7)\,\pcc). Taking into account the total number of bursts, DM, and sky location, \cite{Collaboration:2023aa}
estimated a contamination rate of $R_{\rm cc}=0.92$ for this candidate repeater from the ``silver'' sample ($0.5 \le R_{\rm cc} <
5$); out of 25 ``gold'' and 14 ``silver'' sources, five are expected to be false positive. Given the faint nature of the
latter burst and the resulting uncertainty in its measured parameters, we consider derived properties of FRB\,20200320A
as primary throughout this analysis. The source of this candidate repeater was localized through the CHIME/FRB header
localization pipeline (without baseband data) at right ascension \smash{RA=$42\fdg45_{-0.04}^{+0.02}$} (J2000) and
declination \smash{Dec=$15\fdg84_{-0.06}^{+0.07}$} (J2000), which place the FRB outside the DESI-BGS/LRG survey
footprint. The angular scale of the FRB-galaxy correlation is consistent with the FRB localization error $\theta_f\sim1'$,
corresponding to scales $\ell\sim10^4$ beyond which \clfg\ amplitude is suppressed substantially by the beam
\citep[][]{Rafiei-Ravandi:2020aa, Rafiei-Ravandi:2021aa}. We discuss this atypical
yet robust result in Section~\ref{sec:discussion}.

\begin{table*}[hbt!]
\begin{center}
\begin{tabular}{ccccccccc}
\hline\hline
    \# & WISE ID & RA (J2000) & Dec (J2000) & $W1$ & $W2$ & $B$ & $R$ & $z$ \\
    \hline
    1 & J024925.65+155711.3 & $42\fdg3568758$ & $15\fdg9531491$ & $14.49\pm0.03$ & $14.45\pm0.07$ & $20.8\pm0.1$ & $19.01\pm0.07$ & $0.470$ \\
    2 & J025000.22+154127.0 & $42\fdg5009564$ & $15\fdg6908555$ & $14.33\pm0.03$ & $14.14\pm0.04$ & $20.7\pm0.1$ & $18.82\pm0.07$ & $0.430$ \\
    {\bf 3} & J024948.42+155059.7 & $42\fdg4517860$ & $15\fdg8499300$ & $14.57\pm0.03$ & $14.18\pm0.04$ & $20.35\pm0.09$ & $18.59\pm+0.07$ & {\bf 0.310} \\
    {\bf 4} & J024947.67+155101.3 & $42\fdg4486436$ & $15\fdg8503706$ & $14.76\pm0.03$ & $14.43\pm0.05$ & $20.15\pm0.08$ & $18.46\pm0.07$ & {\bf 0.276} \\
    {\bf 5} & J024949.86+155102.4 & $42\fdg4577628$ & $15\fdg8506816$ & $14.62\pm0.03$ & $14.36\pm0.05$ & $20.8\pm0.1$ & $18.80\pm0.07$ & {\bf 0.367} \\
    {\bf 6} & J024950.98+155111.2 & $42\fdg4624195$ & $15\fdg8531153$ & $14.78\pm0.03$ & $14.44\pm0.05$ & $20.7\pm0.1$ & $18.74\pm0.07$ & {\bf 0.324} \\
    7 & J025015.49+155056.4 & $42\fdg5645709$ & $15\fdg8490154$ & $14.78\pm0.03$ & $14.49\pm0.05$ & $20.03\pm0.08$ & $18.80\pm0.07$ & $0.279$ \\
    8 & J024929.63+154724.0 & $42\fdg3734899$ & $15\fdg7900246$ & $14.50\pm0.03$ & $14.28\pm0.05$ & $19.92\pm0.07$ & $18.35\pm0.06$ & $0.284$ \\
    9 & J024958.24+155535.5 & $42\fdg4927080$ & $15\fdg9265453$ & $14.40\pm0.03$ & $14.13\pm0.04$ & $19.90\pm0.07$ & $18.14\pm0.06$ & $0.287$ \\
    10 &  J024928.10+155511.9 & $42\fdg3671062$ & $15\fdg9199978$ & $15.20\pm0.04$ & $15.05\pm0.08$ & $20.12\pm0.08$ & $19.14\pm0.08$ & $0.282$ \\
\hline\hline
\end{tabular}
    \caption{\label{tab:wisexscos}WISE$\times$SCOS galaxies in the vicinity ($<10'$) of FRB\,20200320A
    (\#3--6, in bold, are $\lesssim 1'$ from the FRB; see Section~\ref{sec:results}). The
    photometric redshift ``$z$'' has an overall uncertainty of $\approx0.033$. The WISE magnitudes ``$W1$''
    (3.4\,\micron) and ``$W2$'' (4.6\,\micron) are in the Vega system, whereas the SCOS magnitudes ``$B$''
    and ``$R$'' are in an AB system \citep[see][]{Bilicki:2016aa}.}
\end{center}
\end{table*}

Besides the positive correlation, we find for the first time statistical evidence for a negative
correlation in the FRB-galaxy cross power spectrum. For instance, we examine
panel (d) of Figure~\ref{fig:clfg}: the ``repeaters (baseband)'' sample correlates negatively
with the DESI-BGS/LRG galaxies on scales $\ell\sim100$. As a case in point, the negative
bandpower centered on $\ell=127.5$ has a local \pvalue\
(not accounting for look-elsewhere factors in $\ell$ or $z$) of 0.002 and $<$\,0.001 for the
DESI-BGS and DESI-LRG surveys, respectively. Computing a global
significance through $\min_{L,z}\SNR_{L,z}$ (see Section~\ref{sec:pipeline-overview}), we
obtain the \pvalues $0.097$ (DESI-BGS, $L=768, z_{\rm min}=0.24$) and $0.006$ (DESI-LRG,
$L=315, z_{\rm max}=0.45$), which fully account for the look-elsewhere effect.

We note that the locally significant negative correlation in the DESI-BGS case vanishes after
accounting for look-elsewhere factors.
In addition, negative correlations from the ``repeaters
(all)'' sample are not statistically significant (\pvalues$\sim0.1$) after accounting for the
look-elsewhere effect. Thus, the only case that requires further consideration
is the global \pvalue$=0.006$ (DESI-LRG, $L=315$, $z_{\rm max}=0.45$).
Furthermore, we note that repeaters and nonrepeaters differ by this negative correlation,
which we discuss in the next section. Finally, we run the cross-correlation pipeline with the
updated parameters $N_{\rm side}=8192$, $\ell_{\rm max}=14000$, and $315 \le L \le 13960$
for CHIME/FRB Catalog 1. We find that all the results presented in this work are consistent
(i.e., no statistical or interpretational discrepancy after considering uncertainty limits throughout)
with the work presented by \cite{Rafiei-Ravandi:2021aa}. In particular, we do not find
any significant difference between results (\pvalues along with their corresponding template
scale and redshift endpoint; see the two rightmost columns of Table~\ref{tab:wisexscos}) from
nonrepeaters and Catalog 1.

\section{Discussion}
\label{sec:discussion}
In this article, we report on a statistically significant correlation (\pvalue$<0.001$) between
a sample of CHIME/FRB repeaters and WISE$\times$SCOS galaxies. We show that the correlation
is dominated by the proximity of FRB\,20200320A (extragalactic $\DM\approx550$\,\pcc) to a group
of four galaxies with mean redshift $z\approx0.32$ (see Figure~\ref{fig:frb20200320a}). In
Table~\ref{tab:wisexscos}, we list all the galaxies from the WISE$\times$SCOS survey in the
vicinity of the FRB. The listed redshift values for the four galaxies mentioned above (galaxies \#3--6)
are consistent within 1.5$\sigma$ ($\approx130\,h^{-1}$Mpc) uncertainty,
owing to photometric redshift errors. Hence, they could all be at the same redshift within the statistical
uncertainty.

Assuming a Navarro--Frenk--White density profile \citep{Navarro:1996gj} and comoving units throughout,
a dark matter halo with mass $M\sim10^{12}~M_\odot$, which is a typical value for galaxy groups
with $\lesssim4$ members \citep{Tully:2015aa}, spans a virial radius of $\sim1'$ ($\sim400$~kpc)
at this redshift on the sky. This picture is consistent with the FRB and galaxy group inhabiting the
same dark matter halo. If we assume that galaxies \#3--10 in Table~\ref{tab:wisexscos} are members
of the same gravitationally bound system, then we estimate the total halo mass to be $M\gtrsim10^{14}~M_\odot$.
Nevertheless, no galaxy clusters have been identified near this sky location at redshift $z\approx0.32$
\citep[see the cluster catalog of][which has a completeness limit of $\approx50\%$ for clusters with mass
$M\approx10^{14}~M_\odot$]{Wen:2018aa}. Additionally, we do not find any Sunyaev-Zel'dovich or X-ray
sources near this group of galaxies \citep{Flesch:2016aa,Collaboration:2016aa,Salvato:2018aa}.
More precise statements on this presumably bound system of galaxies would require
spectroscopic redshifts, which were unavailable at the time of writing.
Therefore, we defer the study of the galaxy group to future work.

If we assume that the FRB originated from redshift $z\approx0.32$, then
the mean IGM contribution to the observed DM is $\Di\approx325\pm180$\,\pcc
\citep{Macquart:2020aa,James:2021ab}. Computing the Galactic contribution based on two
different models, we obtain $\Dg=38.92$\,\pcc \citep[YMW16,][]{Yao:2017aa} and 45.53\,\pcc
\citep[NE2001,][]{Cordes:2002tt} with negligible uncertainties compared with the uncertainties
in $\Di$ and $\DM_{\rm halo}$ (see Section~\ref{sec:introduction}). Subtracting these contributions
from the observed DM, we estimate the FRB host DM to be $\Dh \lesssim 380$\,\pcc,
which also includes any contributions from the host group/cluster electron profile.
As pointed out by \cite{Ibik:2023aa}, most localized FRBs have $\Dh\lesssim200$\,\pcc,
and the FRB host DM distribution, e.g., as a function of redshift, is largely unknown.

Based on the assumptions outlined earlier, we estimate the $\Dh$ contribution from
the host halo encompassing the galaxy group to be $\Dhh \gtrsim 70$\,\pcc \citep{Connor:2022aa}.
In the foreground, we identify the galaxy cluster NSC\,J024958+155217 \citep[Northern Optical Cluster
Survey III,][]{Gal:2009aa} at redshift $z\approx0.12$ along the line of sight. This cluster has $12.0\pm5.6$
member galaxies that constitute a symmetric mass profile with a virial radius of $\approx1$~Mpc
\citep[see Figure~5 of][]{Gal:2009aa}. Assuming a halo mass of $\sim10^{14}~M_\odot$ with an ``ICM''
gas profile \citep{Prochaska:2019ab}, we estimate that NSC\,J024958+155217 contributes
$\gtrsim100$\,\pcc to the $\Di$ of FRB\,20200320A, which is localized to $\approx3\farcm2$
($\approx400$~kpc) from the halo center. However, we highlight that the actual DM contribution from
the host/intervening halo depends strongly on the uncertain distance between the FRB and the halo
center \citep[see, e.g., Figure~11 of][]{Rafiei-Ravandi:2021aa}.

We emphasize that FRB\,20200320A is a ``candidate'' repeater comprised of two
relatively different bursts (see Section~\ref{sec:results}),
making this FRB-galaxy correlation intriguing. For instance, we note that a DM
variation of $\approx13.5$\,\pcc between the two FRBs, separated by 7.5~months, is
large compared with observations of published repeaters that have typically shown
smaller DM variations, e.g., from $<0.15$\,\pcc in 10 months \citep{Nimmo:2023aa}
to $\sim4$\,\pcc in $\sim1.5-3$~yr \citep{Hilmarsson:2021aa,Kumar:2023aa}. Here, the only other
comparator is the ``candidate'' repeating FRB\,20190107B \citep{Collaboration:2023aa} with a DM
variation of $\approx14$\,\pcc in two months.
Furthermore, FRB\,20200320A has a scattering timescale
of 2.46(3)~ms at 600\,MHz \citep[see Figure~13 of][]{Collaboration:2023aa}, $\lesssim$\,1\%
of which could be produced by the Galactic ISM based on the NE2001 and YMW16 models.
Given the FRB redshift, the remaining $\gtrsim$\,99\% of scattering could originate
from the host CGM and circum-burst environment
\citep[][]{Macquart:2013aa,Masui:2015aa,Prochaska:2018aa,Chawla:2022aa,Ocker:2023aa,Sammons:2023aa}.
In sharp contrast, FRB\,20201105A (the second burst from presumably the same source) has a scattering
timescale of $\ll$\,1\,ms at 600\,MHz \citep{Collaboration:2023aa}. These burst-to-burst
variations in DM and scattering further motivate follow-up observations
of the FRB and galaxy group in order to identify and characterize the exact host through interferometry
and spectroscopy at higher angular and frequency resolutions, respectively. With a burst rate of
$3.91^{+14.60}_{-3.78}\times10^{-2}$\,hr$^{-1}$ \citep{Collaboration:2023aa}, any follow-up radio observations
of the FRB (e.g., to verify whether the ``candidate'' is indeed a repeater with two bursts, or two nonrepeaters are
associated with the same galaxy group/cluster) would be feasible only for continuous scans over $\sim$days--month.
In the meantime, spectroscopy of the galaxy group could shed light on our understanding of potential environments
that the FRB source might inhabit.

The association between FRB\,20200320A and a galaxy group is not a typical result from statistical
cross-correlations with large numbers of roughly localized sources; typically, we expect a large fraction
of the sources to contribute statistically to any high-SNR detections (see, e.g., the results on nonrepeaters
in Table~\ref{tab:pvals}). Nonetheless, this positive correlation is robust, since we use an appropriate
template for the one-halo term and account for all look-elsewhere factors throughout the pipeline.
This is the first time that a presumably bound system of galaxies has been discovered using
FRB observations. Indeed, we would obtain the same result if we were to cross-match the FRB sample
with a catalog of galaxy groups/clusters containing the group of four galaxies here. Therefore, we strongly
suggest including catalogs of galaxy groups and clusters in direct host association pipelines even for FRBs
with $\sim1'$ localization precision.

In addition, we find evidence for a negative correlation between our ``repeaters (baseband)" sample
of FRBs and DESI-LRG galaxies. Negative FRB-galaxy correlations were formalized and forecast
analytically by \cite{Rafiei-Ravandi:2020aa}. The formalism was applied for the first time to
real data (CHIME/FRB Catalog 1) by \cite{Rafiei-Ravandi:2021aa}.
In this work, we report the statistical significance of the negative correlation as
\pvalue$=0.006$ ($L=315$, $z_{\rm max}=0.45$), which is interesting yet
marginal. We do not find any such evidence for negative correlations in the analysis
of ``nonrepeaters," suggesting hypothetically that repeaters and nonrepeaters might cluster differently
in dark matter halos over angular scales $\sim0\fdg5$.
However, negative terms in the FRB-galaxy cross power spectrum \clfg\ might arise
only from propagation effects due to intervening plasma \citep{Rafiei-Ravandi:2020aa}.
For instance, a negative DM-completeness term could stem from
DM perturbations caused by electron anisotropy along the line of sight.
These perturbations decrease the probability of detecting FRBs above
a fixed signal-to-noise ratio threshold (i.e., increased DM: fluence preserved,
signal-to-noise ratio reduced). In such a scenario, the observed number density of
FRBs, apparently correlated with galaxies through foreground electrons,
experiences a deficit due to the influence of DM perturbations.
Consequently, this deficit results in the appearance of a
negative contribution to \clfg.
Considering the borderline significance (99.4\% CL, after accounting for
the look-elsewhere effect in $L$ and $z_{\rm max}$) of this negative correlation based on a
relatively small sample size ($N_{\rm FRB}=18$), we defer any additional analysis regarding
it, such as characterizing the instrumental selection function  for the ``repeaters (baseband)''
sample, to future work.

Would we expect the repeaters to show an overall positive correlation with galaxies,
in light of the results on nonrepeaters? This depends on the $\ell$-dependence of the cross-correlations.
The repeater population has larger \clfg\ error bars (see Figure~\ref{fig:clfg}), so if the two FRB populations
had the same $\ell$-dependence, then we would not expect the repeaters to show a statistically significant
cross-correlation. On the other hand, if the repeaters ($\ell$\clfg) peaked at higher $\ell$-values (compared to
nonrepeaters), which is plausible because the repeaters have more precise localization, then we might find a
cross-correlation at higher \Lvalues. Before we did the analysis, either outcome was possible. However, the lack
of an overall positive correlation between the repeating FRBs (excluding FRB\,20200320A) and galaxies is
unsurprising, given the small sample size. The CHIME/FRB baseband system is continuously
localizing repeating and nonrepeating sources, enabling future explorations of the FRB phenomenon through
cross-correlations with large-scale structure. The results in this work will improve with larger FRB catalogs
in the future.

\acknowledgements
We acknowledge that CHIME is located on the traditional, ancestral, and unceded territory of the
Syilx/Okanagan people. We are grateful to the staff of the Dominion Radio Astrophysical Observatory,
which is operated by the National Research Council of Canada.  CHIME is funded by a grant from the Canada
Foundation for Innovation (CFI) 2012 Leading Edge Fund (Project 31170) and by contributions from the
provinces of British Columbia, Qu\'{e}bec, and Ontario. The CHIME/FRB Project is funded by a grant from the
CFI 2015 Innovation Fund (Project 33213) and by contributions from the provinces of British Columbia and
Qu\'{e}bec, and by the Dunlap Institute for Astronomy and Astrophysics at the University of Toronto.
Additional support was provided by the Canadian Institute for Advanced Research (CIFAR), McGill University
and the McGill Space Institute thanks to the Trottier Family Foundation, and the University of British
Columbia. Research at Perimeter Institute is supported by the Government of Canada through Industry Canada
and by the Province of Ontario through the Ministry of Research \& Innovation.
The Dunlap Institute is funded through an endowment established by the David Dunlap family and the
University of Toronto.
\par
K.\,M.\,S. was supported by an NSERC Discovery Grant and a CIFAR fellowship.
Z.\,P. is a Dunlap Fellow.
M.\,B. is a McWilliams fellow and an IAU Gruber fellow.
M.\,D. is supported by a CRC Chair, NSERC Discovery Grant, CIFAR, and by the FRQNT Centre de
Recherche en Astrophysique du Qu\'ebec (CRAQ).
G.\,M.\,E. is supported by an NSERC Discovery Grant (RGPIN-2020-0455), and by a Canadian Statistical
Sciences Institute Collaborative Research Team grant.
B.\,M.\,G. is supported by an NSERC Discovery Grant (RGPIN-2022-03163), and by the Canada Research
Chairs (CRC) program. 
V.\,M.\,K. holds the Lorne Trottier Chair in Astrophysics \& Cosmology, a Distinguished James McGill
Professorship, and receives support from an NSERC Discovery grant (RGPIN 228738-13), from an R.
Howard Webster Foundation Fellowship from CIFAR, and from the FRQNT CRAQ.
C.\,L. is a NASA Hubble Fellowship Program~(NHFP) Einstein Fellow, supported by NASA through the NASA Hubble
Fellowship grant \#HST-HF2-51536.001-A awarded by the Space Telescope Science Institute, which is operated by the
Association of Universities for Research in Astronomy, Incorporated, under NASA contract NAS5-26555.
K.\,W.\,M. holds the Adam J. Burgasser Chair in Astrophysics and is supported by NSF grants (2008031, 2018490).
A.\,P. is funded by the NSERC Canada Graduate Scholarships -- Doctoral program.
A.\,B.\,P. is a Banting Fellow, a McGill Space Institute~(MSI) Fellow, and a Fonds de Recherche du Quebec
-- Nature et Technologies~(FRQNT) postdoctoral fellow.
D.\,C.\,S. is supported by an NSERC Discovery Grant (RGPIN-2021-03985) and by a Canadian Statistical
Sciences Institute (CANSSI) Collaborative Research Team Grant.
This research was enabled in part by support provided by WestGrid (www.westgrid.ca)
and Compute Canada (www.computecanada.ca).
The Photometric Redshifts for the Legacy Surveys (PRLS) catalog used in this article was
produced thanks to funding from the U.S. Department of Energy Office of Science, Office
of High Energy Physics via grant DE-SC0007914.
This research has made use of ``Aladin sky atlas'' developed at CDS, Strasbourg Observatory, France.
The Pan-STARRS1 Surveys (PS1) and the PS1 public science archive have been made possible through
contributions by the Institute for Astronomy, the University of Hawaii, the Pan-STARRS Project Office,
the Max-Planck Society and its participating institutes, the Max Planck Institute for Astronomy, Heidelberg,
and the Max Planck Institute for Extraterrestrial Physics, Garching, The Johns Hopkins University, Durham
University, the University of Edinburgh, the Queen's University Belfast, the Harvard-Smithsonian Center
for Astrophysics, the Las Cumbres Observatory Global Telescope Network Incorporated, the National Central
University of Taiwan, the Space Telescope Science Institute, the National Aeronautics and Space
Administration under grant No. NNX08AR22G issued through the Planetary Science Division of the NASA
Science Mission Directorate, the National Science Foundation grant No. AST-1238877, the University of
Maryland, Eotvos Lorand University (ELTE), the Los Alamos National Laboratory, and the Gordon and Betty
Moore Foundation.

\bibliographystyle{aasjournal}
\bibliography{ch_frbx_rn3.bib}

\end{document}